# Eyes on the Mission: Mixed Methods Assessment of Eye-Tracker-Enabled Interactive Decision Support in a Simulated Unmanned Aerial Vehicle System

Hyun-Gee Jei [a], Mustafa Demir [a], and Farzan Sasangohar [a, *]

[a] Wm Michael Barnes Department of Industrial and Systems Engineering, Texas A&M University

**Competing Interests Statement:** X

**Funding:** This research did not receive any specific grant from funding agencies in the public, commercial, or not-for-profit sectors.

**Author Contributions: Hyun-Gee Jei:** Conceptualization, Methodology, Validation, Formal Analysis, Investigation, Data Curation, Writing-Original Draft, Visualization, Project administration; **Mustafa Demir:** Formal Analysis, Data Curation, Writing-Original Draft, Visualization; **Aishwarya Kanaparthi:** Software; **Jacob Kolman:** Writing-Reviewing&Editing; **Farzan Sasangohar:** Conceptualization, Methodology, Resources, Writing-Reviewing&Editing, Supervision, Project administration

**Manuscript type:** Research Article

**Exact word count: ~5870 (target <5000 or editor inquiry)**

*Corresponding Author. Farzan Sasangohar, 101 Bizzell St, Mail Stop 3131, College Station, TX 77843-3131. Email: sasangohar@tamu.edu; phone: +1-(682)-300-8380

**ABSTRACT**

Supervisors in military command and control (C2) environments face dynamic conditions. Dynamically changing information continuously flows to the supervisors through multiple displays. In this environment, important pieces of information can be overlooked due to the complexity of tasks and environments. This study examined the efficacy of an eye-tracker-based adaptive attention-guided decision support tool (DST) for supervisors in a simulated C2 environment. The DST monitors supervisors' visual attention allocation in real-time and displays visually salient cues if critical changes or events are missed. Twenty-five military students participated in a simulated intelligence task. Results indicated significant performance enhancement when the adaptive DST was present. Eye-tracking analysis also showed that longer, more frequent fixations on critical areas of interest were negatively correlated with performance. Additionally, post-experiment interviews revealed that the adaptive DST was unobtrusive and positively received. These findings underscore the potential of real-time gaze-based interventions to optimize supervisory decision-making. Future research could incorporate AI-driven approaches to better support supervisors in complex task environments.

**Highlights**

- Real-time eye-tracker-based attention-guidance cues enhanced military mission commanders' performance in a high-fidelity command center simulation environment.
- Gaze analysis showed that the real-time eye-tracker-based attention-guidance cues were not only able to improve their performance but also promoted more strategic and efficient visual processing.
- Gaze contingent attention-guided visual cues received positive feedback from the participants for not being obtrusive—88% of participants did not notice the cues, yet still outperformed the control session.

## 1. INTRODUCTION

Mission Commanders (MCs) in modern command, control, communications, computers (C4), intelligence, surveillance, and reconnaissance (ISR) face rapidly evolving dynamic situations that sometimes outweigh their cognitive capacities. ISR-enabled data is vast due to growing capabilities and a network of complex technologies such as satellites and sensors. Furthermore, the advent of AI and enhanced sensing technologies enables information to be transmitted to the command center interfaces in real-time (Fontes & Kamminga, 2023). Consequently, successful decision-making in supervisory control and the maintenance of situation awareness (SA) demand that the MCs perceive and process large amounts of information displayed on various interfaces (Demir et al., 2023; Endsley, 2001; Endsley et al., 2003; S. Humr et al., 2025; S. A. Humr et al., 2023). Such exposure to a large volume of information may result in overload, and information may not be attended to in a timely manner due to non-optimal visual attention allocation. Timely analysis of dynamic information flow across the multimodal sensors has been vital in maintaining SA and effective decision-making (Endsley, 1995), helping forces counter advancements and manage battlefield conditions more efficiently (Rickli & Mantellassi, 2024).

It is well-documented that attention is the bottleneck in human information processing and that attentional resources are scarce (Rogers, 2023; Tsotsos et al., 1995; Wickens, 1976). In modern-day C4 centers, MCs are typically surrounded by multiple mission information displays to maintain SA and inform tactical decisions. They also need to divide their attention for monitoring and communication purposes. Unlike auditory displays, where the signal can be detected omnidirectionally, processing information from visual displays requires optimal gaze. However, given the vast amount of visual information provided in C4, potentially important information may be overlooked and go unnoticed due to inattentional blindness (Mack, 2003; Mack & Rock, 1998; Neisser, 1979; Simons & Chabris, 1999). In addition, frequent interruptions may divert attention away from the mission displays, making MC vulnerable to distractions and attentional issues in general. Another example is change blindness. Given the dynamics of military missions, MCs are also prone to experiencing a phenomenon known as change blindness (Simons & Levin, 1997), in which the subject fails to notice changes in the visual scene due to even a brief visual obstruction (Rensink et al., 1997). Therefore, it is essential to support MCs in complex C4 environments to attend to important information necessary for tasks-at-hand to make timely and sound decisions.

Previous research has investigated attention-guiding methods to support effective attention allocation. In the driving contexts, Calvi et al. (2021) augmented reality (AR) visual and auditory feedback was used to aid drivers in executing safer takeover maneuvers. Additionally, Stahl et al. (2016) applied highlighting color cue to guide attention in vehicle displays, which improved novice drivers' performance. In education,

De Koning et al. (2010), Grant and Spivey (2003), and Roads et al. (2016) implemented attention-grabbing highlights to better guide students in comprehending complex concepts. In aviation and defense, Hopp et al. (2005) and Savick et al. (2008) employed tactile cues via a vibrating vest and belt to guide attention toward important events. In both cases, guiding attention using tactile cues improved performance in a complex task environment. However, tactile cues were also prone to misinterpretation of cue location and meaning (Nonino et al., 2021; Sklar & Sarter, 1999). This can be a critical limitation in high-risk, safety-critical, and dynamic settings. Additionally, Frame et al. (2022) guided users' attention by placing small colored globes above relevant elements, improving recognition of essential elements required to make sense of scenes. Although increasing visual salience can support attentional guidance, some studies found that the high intensity of visually salient features can induce negative emotions (e.g., annoyance) and/or performance decrement (Bahr & Ford, 2011; Lewandowska et al., 2022; Tasse et al., 2016; Ward & Helton, 2022). Moreover, real-time adaptive aid through attention guidance remain underexplored: only one study (Fortmann & Mengeringhausen, 2014) implemented such a system in a military scenario. which was Though not used as a decision support tool via attention-guidance, a similar line of work was conducted by Ratwani and colleagues. They have utilized real-time eye-tracking to predict and prevent postcompletion errors, which can lead to critical consequences in high-risk environments if not treated properly (Ratwani et al., 2008; Ratwani & Trafton, 2011). Breslow et al. also applied similar predictive modeling for workload prediction and error prevention using real-time eye-tracking technology in dynamic scenarios where operators had to manage multiple unmanned aerial vehicles (UAV) (Breslow et al., 2014). However, these forementioned works are different from our work in that those experiments were conducted in a simple environment that did not accurately represent the complex nature of the military command centers. To our knowledge, the application of real-time attention guidance in complex safety-critical work environments remains a research gap.

To address this gap, the goal of this study was to investigate the efficacy of adaptive interfaces by integrating attention-guiding features that respond to gaze behaviors of MCs in a simulated C4 environment. This paper documents the findings from an experiment that was conducted in a high-fidelity facility that mimics C4 control rooms to investigate the efficacy of this tool on improving MCs' decision-making.

## 2. MATERIAL AND METHODS

### 2.1 Simulation Testbed

The simulated task environment emulated a ground force protection mission, specifically safeguarding a critical political convoy navigating through hostile territory. In this environment, a team composed of a human MC and three computer-simulated operators controlled three semi-autonomous UAVs to scan for

short-, medium-, and long-range missiles. The team also had to coordinate with an external air strike team to destroy any identified threats to the convoy. Additionally, the MC had access to various assets to complete the convoy mission, such as UAVs, strike teams, and Airborne Warning and Control Systems (E-3 AWACS).

The simulation testbed running this scenario consisted of three 70-inch interactive displays (Dell C7017T) and a 55-inch touchscreen tabletop display, visualized in Figure 1. Each display provided various types of mission-critical information. The Mission Status Display (Figure 1, top-left) displayed operator performance metrics, UAVs' status, and communication status. It also showed the history of messages exchanged between the MC, UAV operators, and the airstrike team. When UAVs took over 30 seconds to identify a discovered threat, the MC could request information and assist in threat identification via the Remote Assistance Display (Figure 1, top-right). The Map Display (Figure 1, top-center) served as the main display and provided a geospatial map of the mission, three UAV operators' area of interest, the current location of the convoy, discovered threats and their ranges, areas scanned by the UAVs, convoy health status, communication status, elapsed mission time, and the timeline to show strike schedule, potential and known threats, and when certain UAVs were destroyed. Finally, the Mission Commander Display (Figure 1, bottom-center) was provided on a tabletop and allowed MCs to make decisions and issue orders. Throughout the mission, the MC must stay alert to monitor multiple large displays to track the mission status and continuously make decisions regarding the convoy's health status, each UAV operator's decision-making performance, and UAV movements. Successful completion of the mission not only involved monitoring the progress of the convoy and UAVs but also required the MC to take timely actions such as holding or resuming the convoy, re-assigning UAVs, and rerouting the convoy to a safer route.

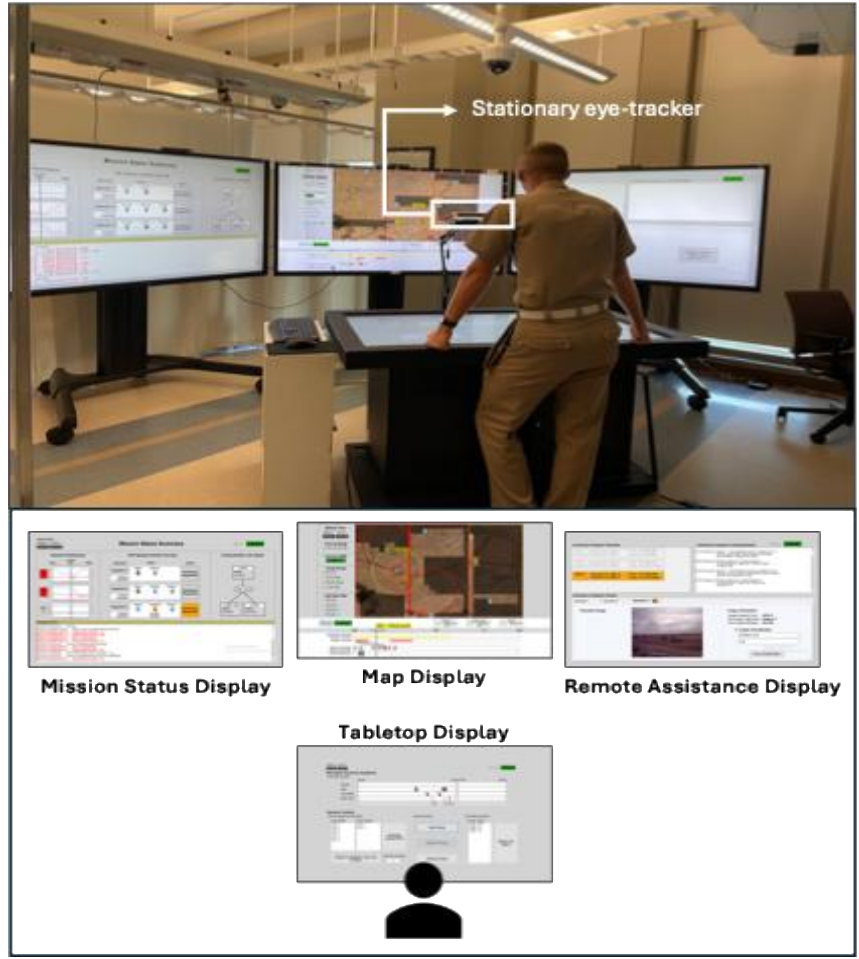

**Figure 1.** Simulation Testbed with its. The mission commander (participant) communicates with the simulated UAV operators and E-3 AWACS to oversee the ground force protection mission.

## 2.2 Eye-Tracker-Based Adaptive DST

Throughout the simulation, a stationary eye-tracker (Tobii Pro Nano; sampling rate 60Hz with 0.3° accuracy) was installed in front of the main map display (Tobii AB, 2023) to continuously monitor the MCs' gaze behavior. The eye-tracking data served two purposes: providing a real-time data feed to support the adaptive attention-guided DST, and enabling post hoc analysis of MC's visual attention patterns to examine how visual attention allocation influenced decision-making and overall performance. All eye-tracking data was recorded and exported using the Tobii Pro Lab software (version 1.123).

For real-time attention guidance, if the MC failed to identify crucial mission changes (e.g., UAVs being destroyed or the convoy entering a threat range) within 3 seconds of the changes, the system presented a combination of visual salient cues, including a warning message, color changes, and blinking to highlight the corresponding area of interest (AOI; Figure 2). For example, when the convoy is attacked or UAVs are destroyed by enemy artillery, MCs receive notification messages right below the map (Figure 2, box 1). Additionally, when the communication link is disconnected from the convoy, strike team, or E-3 AWACS, the status color under the map would change to red to attract MC's visual attention (Figure 2, box 2). Lastly, a red box also blinks in the timeline area when there is a late strike (Figure 2, box 3). However, if MC's visual attention was already allocated to the relevant AOI, no attention-guiding feature was displayed to prevent information and visual overload. The location of the salient cues were strategically placed just above the timeline, which is one of the most attended areas of interest (AOI), applying the proximity compatibility principle (Wickens & Carswell, 1995). This placement was intended to minimize the effort required for participants to notice and shift attention to the cues (Wickens, 2015). The key areas of interest (AOI) and adaptive DST features are described in Table 1.

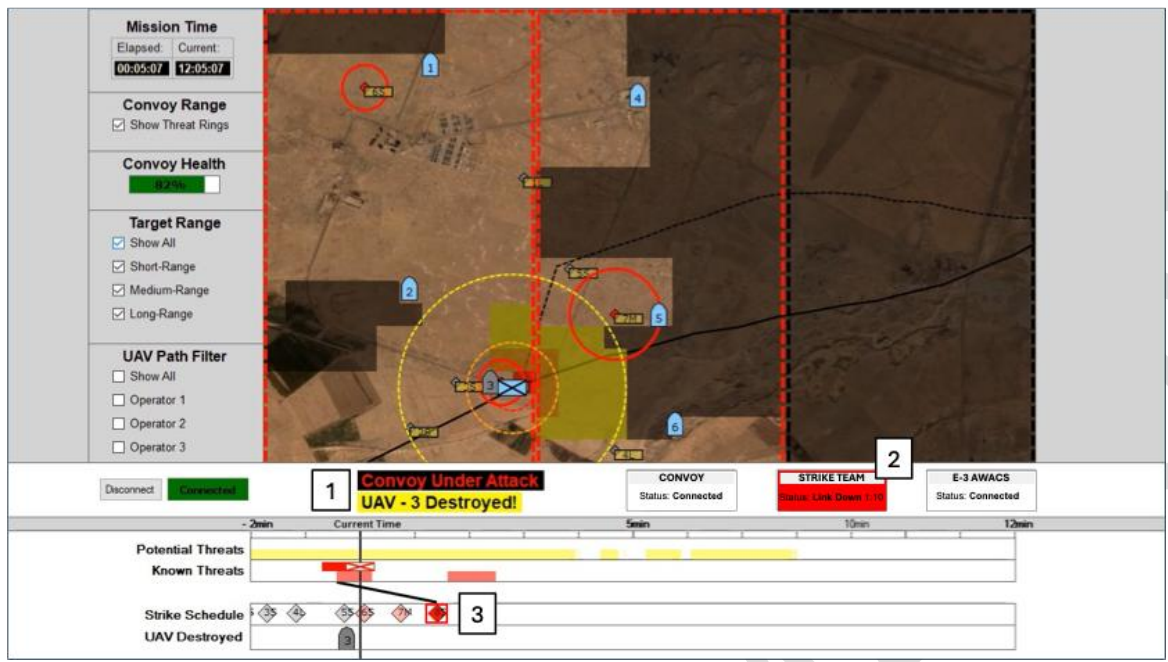

**Figure 2.** Examples of visually salient attention-guided adaptive DST. The messages above the timeline and the red flashing box around the strike schedule in the timeline are examples of adaptive DST.

**Table 1.** Definitions of key AOIs in the map display and key adaptive features.

| AOI | AOI: Definition | Adaptive Feature(s) |
|---|---|---|
| Timeline | The area on the map displays where the interruption recovery assistance timeline is shown. | Flashing red box around late strike targets. |
| Map | Part of the map display shows the map of the operation area divided into three regions. Each region is managed by a virtual UAV operator (3 UAV operators in total). | No adaptive features. However, the adaptive features are activated on other parts of the display when participants fail to notice important changes in the map area. |
| Convoy Health | A progress bar on the map displays the 'life' of the convoy until the mission is unsuccessful. | No adaptive features. However, this AOI is the only real-time performance indicator available to MCs, potentially influencing their decision-making. |
| Messages | The area on the map displays messages regarding the convoy and UAV status. | Messages are adaptively displayed when the MC does not recognize the UAV status. |
| Comm. Status | The map's area shows the communication connection status between the mission commander, convoy, E-3 AWACS, and the strike team. | Blinking red box. |

## 2.3 Design and Hypotheses

This study employed a mixed 2x2x2 design: The within-subjects factors were Condition (Control vs. Adaptive DST) and Scenario (Scenario 1 vs Scenario 2). The between-subjects factor was Trial Order (whether participants experienced the Control or Intervention condition first). The order of condition exposure was counterbalanced. This design allowed us to examine the main effects of Condition and Scenario, as well as whether the order of exposure influenced outcomes through potential learning effect. Each participant completed two scenarios during the experimental session: one under the control condition, where no adaptive DST was shown on the main map display; and one under the intervention condition, where the participants had access to the adaptive DST. The two scenarios were designed to be equivalent in difficulty. We hypothesized (H1) that the MCs' decision-making performance would improve in the intervention condition compared to the control condition.

## 2.4 Sample Size and Participants

A power analysis was conducted using G*Power3 (Faul et al., 2007) to test the difference between the means of 2-condition (within-subjects) by 2-scenario (within-subjects) by 2-trial-orders (between-subjects) using an F-test, with a medium effect size ($\eta$p2 = 0.08; Cohen, 2013), and an alpha ($\alpha$) of 0.05. The priori power analysis was based on the within-subjects portion (Condition x Scenario), since the within-subjects contrasts are the primary effects of interest, while the between-subjects factor (Trial Order) was included primarily as a blocking variable to account for potential learning effects. According to the result, a total sample of 25 participants was required to achieve a power of 0.80.

To rely on military training and education perspectives (Fletcher, 2009), 25 volunteer participants from the Reserve Officers' Training Corps (ROTC) and the Veteran Student Association were recruited at a large Southwestern University in the United States of America. Participants were required to be 18 years or older, be fluent in English, self-report to have normal or corrected-to-normal hearing and vision (e.g., colorblind participants were excluded from this study), and be comfortable using a standard computer mouse and keyboard. Participants were compensated $50 for 2.5 hours spent during the study. Results from two participants' Intervention condition trial were excluded because the eye-tracker could not be properly calibrated, resulting in a total of N = 23 participants with complete data and 2 partial. All retained participants completed both conditions (Control and Adaptive DST) across both scenarios; Trial order (Control-first vs. Intervention-first) was counterbalanced. Participants were generally younger adults (MeanAge = 22.2, SDAge = 3.57), skewed male (88% identified as male), and served in the military for 2-4 years. Of the 25 participants, 13 served or were contracted with the Air Force, three with the Army, one with the Navy, and one with the Coast Guard. Six participants were not affiliated with service branches but were part of the university's Corps of Cadets' drill and ceremonies team. This study was approved by the university's Institutional Review Board (IRB2022-0997D). Participants gave written informed consent.

## 2.5 Procedure

All participants took part in a single-session experiment that lasted approximately 2.5-3 hours. First, participants read through 20-25 minutes of tutorial slides on the experimental procedure, mission summary, goals, and descriptions of the functions of the different displays used during the experiment. Participants were then shown six screenshots of the mission in various phases. Participants were asked to identify the best course of action in the scenario and explain the reasons behind their decision. This mini quiz was given to ensure that participants understood all the materials in the tutorial slides. Participants were also briefed on how their overall performance would be measured. Participants then moved to the experimental environment to complete a training scenario that emulated the testing condition. As they interacted with the testbed, the training trial also served as the benchmark test to ensure they had sufficient knowledge about the mission and the testbed before starting the real trials. The participants had to have 25% or more convoy health remaining, have utilized functions such as holding/releasing convoys, reassigning UAVs, and using E-3 AWACS to pass the benchmark test, and then proceed to the actual experiment trials. The training trial was repeated until the participants met the set requirements. The tutorial session took 45-60 minutes overall.

Participants carried out both a control and an intervention condition. The order of the conditions and scenarios was counterbalanced across participants. The two conditions and the order of the conditions were blinded to the participants to prevent any biases. After completing the first trial, a 5-minute break was given before starting the second trial. During each trial, the participants were taken away from the main task twice at pre-determined timings to carry out interruption tasks. The interruption tasks were both cognitively and visually engaging (i.e., mission planning based on geographical characteristics and location-finding on a map when given geographical coordinates). These interruption tasks were designed to shift their focus away from the main task and emulate a dynamic mission environment of the real world, where MCs are often interrupted. When the convoy reached a designated location, participants were instructed to return to the main task and were required to make operational decisions. As described in Table 2, each scenario included two decision points: one simple and one complex decision. The order of the decision type varied. Scenario 1 presented a simple decision after the first interruption and a complex decision after the second interruption. The order was reversed in Scenario 2. All participants were brought back to the main task at the identical location in each scenario. That way, everyone encountered the same operational decision points. Overall, the experimental trials took about 40-50 minutes.

**Table 2.** Decision types and available actions per decision type.

| Decision Type | Available Actions |
|---|---|
| Simple | Only one possible course of action (COA) to address the ongoing situation (e.g., Convoy is about to enter a threat zone; therefore, MC must hold the convoy). |
| Complex | Multiple COAs area available, but only one COA is optimal to address the ongoing situation (e.g., Convoy is about to enter the unscanned main path, but the alternate path is already scanned and cleared by the UAVs. MC can request E-3 AWACS, re-route, or hold convoy. Re-routing would yield the best outcome for the convoy). |

The Tobii Pro Nano eye-tracker was used throughout the two trials to keep track of participants' visual attention allocation in real-time. Since the accuracy of the eye-tracker calibration was crucial to providing a precise DST, the eye-tracker was calibrated at the beginning of each trial using the nine-point calibration and validation procedure embedded in the Tobii Pro Lab software (Tobii AB, 2019). Upon completion of both trials, participants were asked to fill out a demographics form and were interviewed about their experience during the experiment. The interview was conducted in a semi-structured format with seven pre-designated questions (see Appendix). Some follow-up questions were asked when deemed necessary to get a better idea of how participants gathered information and their decision-making process. The interview lasted around 10-20 minutes.

## 2.6 Measures

*Performance.* Performance was evaluated using a composite scoring system (0-100 points), designed to reflect the multifaceted nature of operational success. Recognizing that multiple factors influence real-life mission outcomes, this method incorporated three critical elements directly impacting the mission's results. This composite scoring method was taken from the military's common performance measurement, Measure of Performance (MOP). According to a U.S. military joint publication, MOP is "a criterion used to assess friendly actions that are tied to measuring task accomplishment" (Joint Staff J-7, 2011). Three quantifiable MOP elements were chosen to calculate overall performance and weighted to reflect mission priorities conveyed in the participant tutorial: convoy health left at the end of the mission (50%), time to complete each trial (35%), and number of late strikes reported (15%). At the beginning of each trial, the convoy's life bar started at 100%. As the mission progressed, the convoy's health decreased if it was under attack from enemy artillery or if it remained idle for too long. The life bar was continuously displayed on the map, enabling participants to monitor the convoy's health status. The highest weight was given to convoy health, as protecting the convoy was the primary goal of the mission. Therefore, the total health percentage left at the end of the mission was used to calculate the MOP score. Similarly, the completion time and the number of late strikes reported were also captured by the simulation testbed from the start to the end of the simulation. Participants reported late strikes if the strike team's neutralization of the enemy asset happened after the convoy had already arrived in the threat area and would have been attacked if not held just before entering the threat area.

*Eye tracking measures.* The specific areas of interest (AOIs) are highlighted in an annotated map display in Figure 3. These areas are identified based on the significance of the task, which is determined by the amount of information obtained from each area and the level of attention it receives. Table 1 summarizes these AOIs, including their adaptive features for four eye tracking measures: (1) Total Duration of Fixation (TDF): the total time a participant's gaze remains fixed on a particular area within the visual environment; (2) Average Duration of Fixation (ADF): the average time length of time the eyes fixed on a specific point; (3) Number of Fixations (NF): The total number of individual fixations within a particular area of interest within the visual environment; (4) Average Pupil Diameter (APD): the average amount of light entering the eye and a reflection of the brain's state and other factors (Krejtz et al., 2018; Van Der Wel & Van Steenbergen, 2018; Van Gerven et al., 2004).

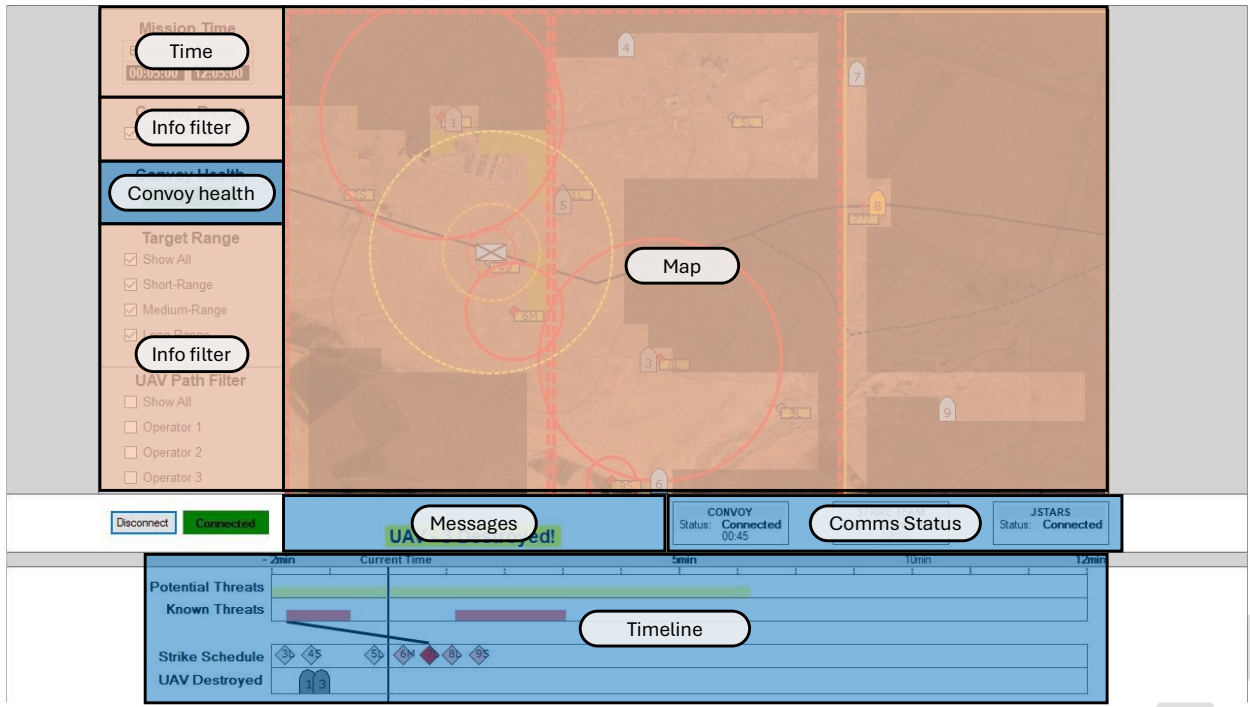

**Figure 3.** AOIs on the annotated map display. Among the eight AOIs, convoy health, messages, Communication (comms) status, and timeline were the key AOIs concerning the mission.

*MC's Decision-Making (time to first decision).* Time to make the first decision after experiencing interruptions, also called resumption lag (Trafton et al., 2005), is essential since decision-making capabilities deteriorate after returning to the main task from interruptions (Monk et al., 2008). Therefore, this measure was added to investigate whether the adaptive attention-guided DST could aid MCs in making more timely decisions after interruptions.

## 2.7 Analysis Methods

A 2 (Condition: adaptive DST or no attention guidance support) × 2 (Scenario: simple or complex decision) × 2 (Order of Trials: between-subjects) split-plot ANOVA was conducted to examine how MOP scores differed across conditions, with participants treated as a random intercept. Including Trial Order allowed us to evaluate whether exposure sequence influenced the effectiveness of the adaptive DST, so accounting for potential learning effect. We assessed whether the normality assumption for the MOP was met using Shapiro-Wilk tests. The Shapiro-Wilk test indicates that the MOP data followed a normal distribution, W(47) = 0.98, p = .452. A Compound Symmetry covariance structure was chosen, assuming equal variances and covariances between repeated measures. The model was also estimated using the maximum likelihood method. Estimated marginal means were calculated for the interaction between condition, scenario, and order, with pairwise comparisons conducted using the least significant difference method.

The normality of the eye-tracking response variables was assessed using the Shapiro-Wilk test, which revealed significant deviations from normality for all the eye-tracking measures: TDF (W = 0.79, p < .001); ADF (W = 0.97, p < .001); NF (W = 0.82, p < .001); and APD (W = 0.99, p = .016). Given the significant normality violations and moderate to high skewness in multiple variables, non-parametric Wilcoxon signed-rank tests were conducted to compare control and intervention conditions for each AOI and metric. Analyses were restricted to participants with complete paired data for both conditions.

To analyze the decision-making time, a normality test was first conducted for the latency variable using the Shapiro-Wilk test. Prior to analysis, the latency data exhibited significant positive skewness (skewness = 3.19) and deviated from normality, as indicated by the Shapiro-Wilk test, W(96) = 0.61, p < .001. Because of the normality assumption violation, we conducted a 2×2 within-subjects design aligned rank transform Analysis of Variance (ANOVA) to examine the effects of condition (control vs. intervention) and decision type (simple vs. complex) on latency.

## 3. RESULTS

### 3.1. Condition and Order Effects on Measures of Performance

The ANOVA results, summarized in Table 3, showed a significant interaction effect between the condition and trial and a significant main effect of the condition (p = .035 and p = .013, respectively). While overall participants performed significantly better during the intervention condition trial compared to the control condition trial, such an effect was observed only when they completed the control condition first, followed by the intervention condition (p < .001). However, their performance was similar when the trial order was reversed (p = .770; Figure 4, right). To evaluate whether the overall patterns was driven by outliers, we also examined boxplots. No extreme cases were identified, and the pattern of results remained unchanged when borderline cases were removed. Yet, variability was slightly larger in the Intervention condition; it appears to reflect natural performance spread rather than undue influence of individual observations.

**Table 3.** Split-plot ANOVA results to predict MOP.

| Source: | $df_{numerator}$ | $df_{denominator}$ | $F$ | $P\ value$ |
|---|---|---|---|---|
| **Condition** | 1 | 22.85 | 7.19 | **0.013** |
| **Trial Order** | 1 | 22.85 | 0.38 | 0.543 |
| **Scenario** | 1 | 22.85 | 0.07 | 0.797 |
| **Condition by Trial Order** | 1 | 24.97 | 5.04 | **0.035** |
| **Condition by Scenario** | 1 | 24.97 | 0.16 | 0.689 |
| **Scenario by Trial Order** | 1 | 24.97 | 0.10 | 0.076 |
| **Condition by Scenario by Trial Order** | 1 | 22.85 | 3.46 | 0.751 |

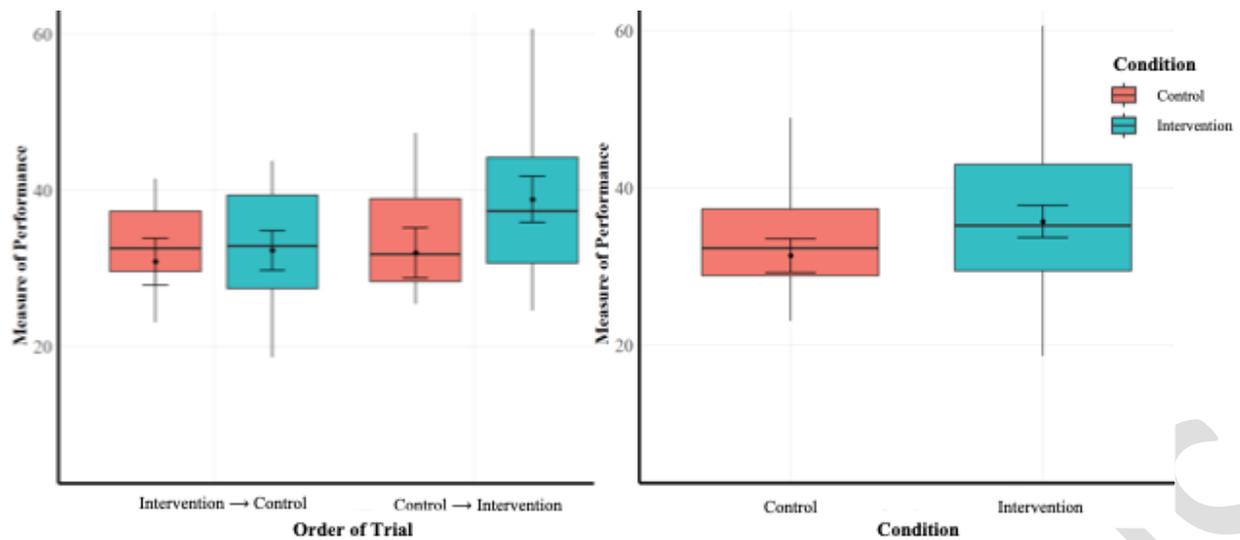

**Figure 4.** Measure of Performance (MOP) score across the conditions and the trial orders (left), and across the conditions (right). Error bars indicate 95% Confidence Interval.

## 3.2. Eye Tracking Measures across Areas of Interest

Table 4 summarizes the Wilcoxon signed-rank tests. For each AOI and measure, the number of paired observations (n), test statistic (V), p-value, standardized z-score, and effect size (r) are reported. Median and interquartile range (IQR) values are provided for both conditions. Significant differences were found in the Convoy Health for ADF (p = .012, r = .52) and TDF (p = .046, r = .41), both of which had longer durations in the control condition than in the intervention (Figure 5). Although not statistically significant, TDF on the messages AOI showed a trend-level difference between conditions (p = .073, r = .37), with participants in the control condition spending more time fixating on this region (Figure 5).

**Table 4.** Wilcoxon Signed-Rank Test results comparing control and intervention conditions across areas of interest and eye-tracking measures.

| AOI | Measure | $n$ | $V$ | $p$ | $z$ | $r$ | Control | | Intervention | |
|---|---|---|---|---|---|---|---|---|---|---|
| | | | | | | | Median | IQR | Median | IQR |
| **Comm. Status** | **ADF** | 23 | 126.5 | 0.738 | -0.33 | 0.07 | 242.00 | 55.5 | 252 | 71 |
| | **APD** | 23 | 152.0 | 0.687 | 0.40 | 0.08 | 3.13 | 0.32 | 3.09 | 0.39 |
| | **NF** | 24 | 135.0 | 0.939 | 0.08 | 0.02 | 14.00 | 11.00 | 15 | 15.75 |
| | **TDF** | 24 | 141.0 | 0.812 | -0.24 | 0.05 | 4122 | 2508.25 | 3372.5 | 4326 |
| **Convoy Health** | **ADF** | 24 | 237.0 | **0.012** | 2.53 | 0.52 | 335.50 | 93.50 | 301 | 53.5 |
| | **APD** | 24 | 113.0 | 0.303 | -1.03 | 0.21 | 3.12 | 0.37 | 3.12 | 0.45 |
| | **NF** | 24 | 184.5 | 0.331 | 0.97 | 0.20 | 54.50 | 25.50 | 45.5 | 44.75 |
| | **TDF** | 24 | 220.0 | **0.046** | 2.00 | 0.41 | 18969.5 | 13614.75 | 13174 | 11241.5 |
| **Messages** | **ADF** | 24 | 187.5 | 0.290 | 1.06 | 0.22 | 194.50 | 60.00 | 181.5 | 36.25 |
| | **APD** | 24 | 146.0 | 0.922 | 0.10 | 0.02 | 3.20 | 0.37 | 3.17 | 0.33 |
| | **NF** | 24 | 196.5 | 0.189 | 1.31 | 0.27 | 32.50 | 16.75 | 25 | 12.5 |
| | **TDF** | 24 | 197.5 | **0.073** | 1.79 | 0.37 | 6087 | 4688 | 4721.5 | 2652.5 |
| **Timeline** | **ADF** | 24 | 145.0 | 0.900 | 0.13 | 0.03 | 348 | 89.00 | 354.5 | 98.5 |
| | **APD** | 24 | 159.0 | 0.812 | -0.24 | 0.05 | 3.12 | 0.41 | 3.11 | 0.35 |
| | **NF** | 24 | 175.0 | 0.484 | 0.70 | 0.14 | 267.5 | 144.5 | 253.5 | 97.5 |
| | **TDF** | 24 | 162.0 | 0.747 | 0.32 | 0.07 | 97587.5 | 57945.5 | 81515.5 | 69155 |

**Note.** ADF: average duration of fixation; AOI: area of interest; APD: average pupil diameter; NF: number of fixations; TDF: total duration of fixation.

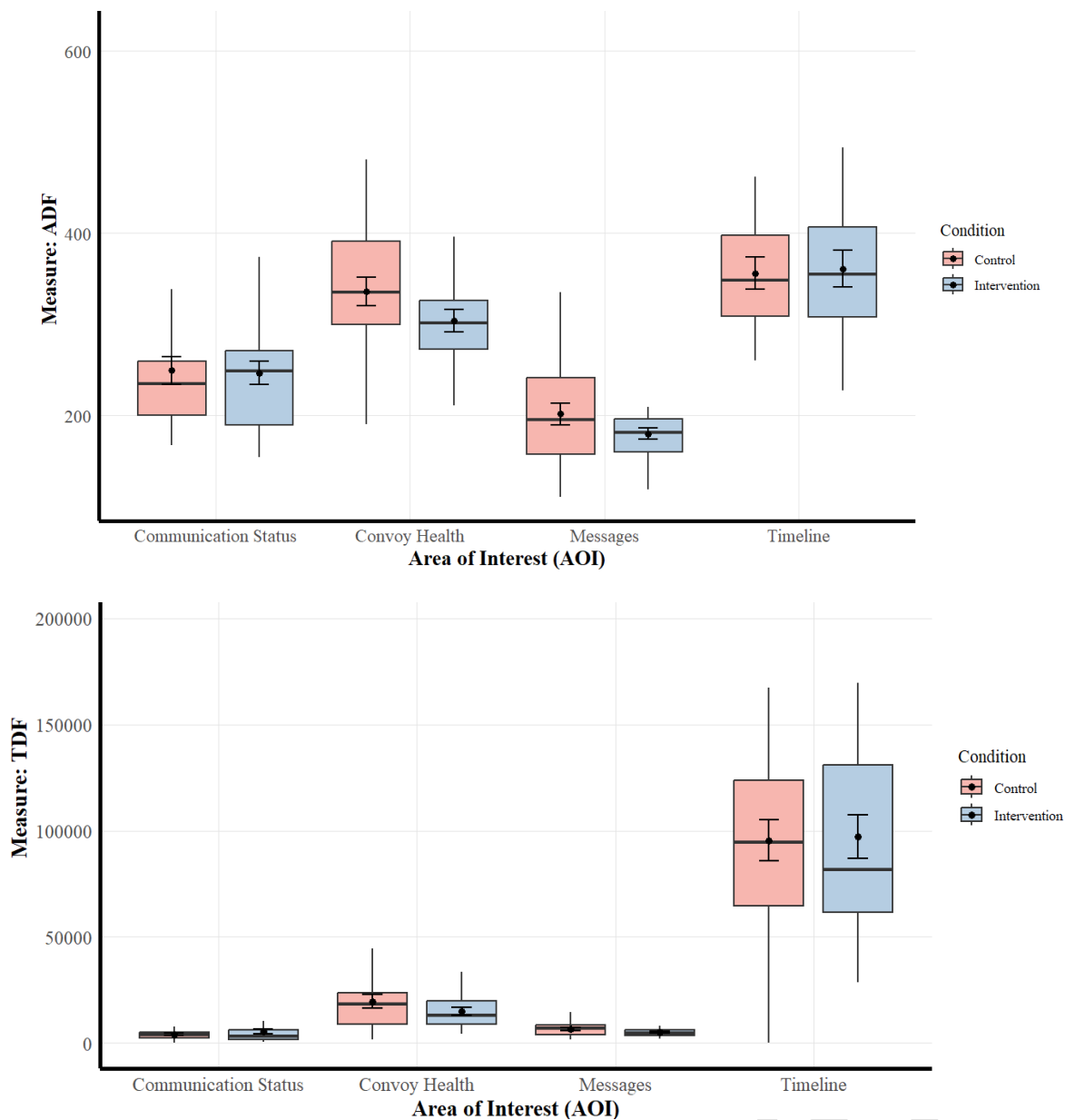

**Figure 5.** Average Duration of Fixation (top) and Total Duration of Fixation (bottom) between the conditions across the selected Areas of Interest. Error bars represent 95% CI.

### 3.3. Decision-Making: Time to Make the First Decision

The ANOVA results revealed a significant main effect of decision type (F(1, 68.49) = 4.38, p = .040). However, there was no significant main effect of condition or interaction between condition and decision type. In particular, participants took longer to respond during complex decisions (Control: Median = 11 s, IQR = 16; Intervention: Median = 7 s, IQR = 10.5) compared to simple decisions (Control: Median = 4 s, IQR = 13; Intervention: Median = 5 s, IQR = 9; see Figure 7). Follow-up pairwise comparisons using aligned rank-transformed data showed that for simple decisions, there was no significant difference between Control and Intervention conditions, t(70) = 1.00, p = .321. Similarly, for complex decisions, resumption lag did not significantly differ between conditions, t(70) = 0.48, p = .631.

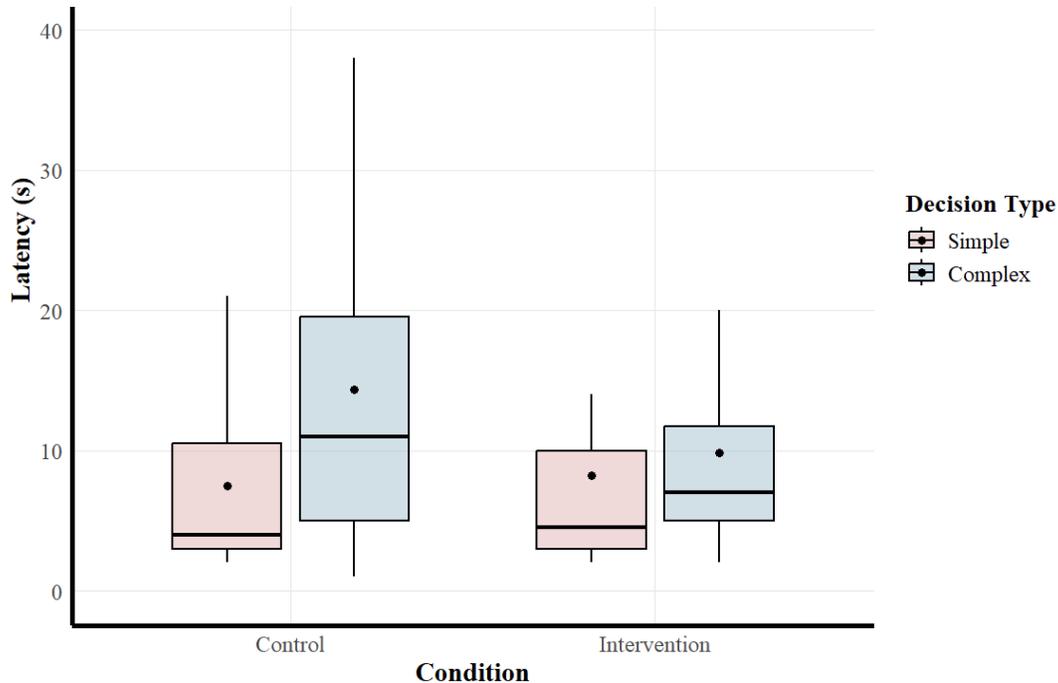

**Figure 6.** Time to make the first decision after experiencing interruptions by decision type (complex vs. simple) and experiment condition (control vs. intervention). Error bars represent 95% CI.

### 3.4. Post-Experiment Interview Results

Interviews were conducted with participants to gather insights into their experience with the testbed and the experimental procedures. The interviews revealed three primary findings regarding the experiment: (1) all participants reported a positive experience with the testbed and generally enjoyed the experiment process despite its lengthy duration (2.5 hours); (2) the adaptive DST features were found to enhance decision-making by encouraging participants to plan ahead rather than react to ongoing situations; and (3) the adaptive DST features were subtle enough that, while some participants noticed them during the intervention, most did not attribute differences between the two trials to these features.

When asked about their overall experience, all participants gave positive feedback, with some describing the experiment as game-like. Regarding the adaptive DST, one participant expressed feeling more secure while managing the mission due to its features.

*"I think when I was playing the game [the experiment], I was trying to have continual situational awareness…. [the messages] made me feel a little more secure and aware." – P21*

Participants mentioned that the adaptive attention-guided DST features helped them plan and take timely actions, especially for UAV surveillance, to protect the convoy as it progressed through the hostile region.

*"Yeah, so when I would see the UAV destroyed [message], I'd look on the map to see which one [got destroyed] and see if it's in a critical area. If it wasn't, then I could just leave it alone. But if it was, then I'd kind of have to time UAV reassignment…. So I don't have to wait for other UAVs to come in later. It's already on the scene." – P11*

*"Oh yeah, I remember [seeing adaptive messages] … when I made decisions, 'Okay then if that's the case, if the UAV is down, then I have to pull other UAVs from the different places'. Yeah, that's when I started to get more alert…it was definitely helpful…without those, then I might have been a little bit late [in re-assigning UAVs]." – P17*

## 4. DISCUSSION

Previous studies conducted using the same testbed focused on providing cognitive support to MCs after they had experienced interruptions (Sasangohar et al., 2014; Scott et al., 2009). These studies employed an interactive timeline where MCs could quickly regain SA based on events marked on the timeline. However, as mission environments grow in complexity, there is an increasing need for dynamic and continuous support, not just post-interruption recovery. Therefore, in this study, a real-time eye-tracking-based attention-guided adaptive DST was developed to enhance supervisory-level decision-making performance.

Our findings suggest that real-time eye-tracking-based attention guidance improves the overall performance. This is consistent with previous research demonstrating that real-time adaptive visual cues can enhance monitoring behavior and SA—factors that can significantly impact performance (Fortmann & Lüdtke, 2013; Fortmann & Mengeringhausen, 2014). Whereas prior studies were conducted in a relatively simple environment, our study demonstrated the efficacy of such support in more complex task environments. Additionally, the improved performance may also be attributed to enhanced attentional selectivity. Rather than scanning all available information for potential threats and events, MCs could focus on prioritized information. This aligns with findings by Frame et al. (2022), who showed that users performed better when the most relevant scene elements were visually emphasized. Their study compared two visual attention aids for essential element identification in military intelligence. One highlighted all humans and vehicles, while the other selectively highlighted only the most relevant elements, including less obvious cues such as weapons and gas canisters. Despite highlighting fewer items overall, the selective tool led to better identification. Ultimately, Frame et al. suggest that more selective attention guidance to relevant elements rather than all potential elements could enhance sensemaking and performance (Frame et al., 2022). These results once again highlight that effective decision support in complex environments depends not on providing more information, but on delivering better-guided information to optimize attention allocation.

An interesting finding was that participants who experienced the intervention condition during their second trial showed significant improvement in overall performance scores. Although participants who received the intervention in their first trial also had slightly better performance scores compared to the control, this difference was not statistically significant. The observed Condition x Trial Order interaction suggests that participants benefited most from the adaptive DST when it followed an initial exposure to the Control condition. One possible explanation is that those who completed the control condition first and then received the intervention benefited from both the learning effects of the initial trial and the adaptive DST. This pattern could reflect a learning effect, where participants developed baseline familiarity during the control condition and then leveraged DST more effectively in the second trial. While this finding underscores the potential of DST in supporting trained operators, it also limits the generalizability of the effect to first-time or novice users. More work is warranted to validate these findings using a between-subjects design to rule out the influence of the learning effect.

Significant differences were found in both average duration of fixations and total duration of fixations in the convoy health area, both being longer in the control condition than in the intervention. The intervention targeted participants' attention allocation, significantly reducing both ADF and TDF in the convoy health area. This may indicate that the participants had to check the convoy's health more frequently and for a longer period during the control condition, as the health decreased more frequently and quickly due to holding the convoy for a longer time and being attacked more often by hostile forces, all of which are indications of poor performance. No other AOIs showed significant differences, suggesting the intervention's influence was specific rather than broad across the interface. Although not statistically significant, TDF on the Messages AOI showed a trend-level difference between conditions, with participants in the Control condition spending more time fixating on this region. This moderate effect may suggest that the intervention helped participants process message content more efficiently by guiding the participants' attention to other more relevant areas on the display. However, further investigation is needed.

Participants in our study did not report annoyance or discomfort with the salient features of the adaptive interface, indicating an unobtrusive user experience while enhancing their performance. This is opposed to some studies, where even though highly salient visual features helped direct attention in time, participants experienced high annoyance at those features (Bahr & Ford, 2011; Lewandowska et al., 2022; Takahashi, 2017; Tasse et al., 2016; Ward & Helton, 2022). In fact, the adaptive attention-guiding cues were so subtle that 88% of the participants reported not noticing them. Some improvement in MCs' performance could be explained by the findings from Vlassova et al., which suggests that unconscious (i.e., unattended) information can still influence people's decision-making (Vlassova et al., 2014). However, more studies are needed to confirm the direct relationship between improved performance and unconscious information in complex settings like ours. This is encouraging as it suggests that adaptive interfaces may support supervisors in complex environments without inducing negative emotional responses, which is especially important in complex environments where human operators can experience high stress.

**Limitations**

This study had several limitations. First, the study had a relatively low sample size and relied mostly on a single-site recruitment strategy, which primarily involved students with military training. Second, although the experimental environment closely mimicked real-world military command centers, it remained a simulation. Future research is warranted to evaluate the efficacy of such tools in the real-world C4 environments with real mission commanders. Third, while MOP is commonly used in military mission evaluations, it is not a validated metric of human performance in the human factors field. A more rigorous validation process is needed to be accepted as a reliable performance measure in human factors research. Lastly, even though the participants showed improvement in performance with the adaptive DST present, the area where the adaptive attention-guiding DST is shown was relatively a small portion compared to the whole viewing field. The location of the adaptive DST was strategically placed to increase the probability of noticing. However, attention-guidance would be ineffective if the salient features are displayed outside of the MC's field of view. Therefore, additional study is currently underway to test the efficacy of adding auditory saliency in better guiding attention, even when MCs are looking completely away from the main display where the visual adaptive DST is shown.

**5. CONCLUSIONS**

This study demonstrated the promise of adaptive DST in enhancing decision-making for MCs in complex environments. However, further investigation is necessary to evaluate the effectiveness of multimodal attention-guidance systems that can dynamically direct attention to the most critical information during high-pressure, time-sensitive operations, as well as proactive decision support tools that can continuously monitor real-time data to help with effective decision-making. Such innovations could dramatically enhance mission success and operational efficiency in the future.

**ACKNOWLEDGEMENT**

The authors thank Jacob M. Kolman, MA, ISMPP CMPP™, senior technologist at Texas A&M University and senior scientific writer at Houston Methodist Academic Institute, for reviewing the language and clarity of an earlier version of this manuscript. Additionally, the authors thank the many undergraduate and graduate students (Christian Perez, Cody Stanford, Michael Lin, Kenny Fisher, Gary Bittick, Omar Hernandez, Jacqueline Hernandez, and Aishwariya Kanaparthi) who have provided tremendous help in designing and executing the experiment.

**REFERENCES**

Bahr, G. S., & Ford, R. A. (2011). How and why pop-ups don't work: Pop-up prompted eye movements, user affect and decision making. Computers in Human Behavior, 27(2), 776–783. https://doi.org/10.1016/j.chb.2010.10.030

Breslow, L. A., Gartenberg, D., McCurry, J. M., & Trafton, J. G. (2014). Dynamic operator overload: A model for predicting workload during supervisory control. IEEE Transactions on Human-Machine Systems, 44(1), 30–40. https://doi.org/10.1109/TSMC.2013.2293317

Calvi, A., D'Amico, F., Ferrante, C., & Ciampoli, L. B. (2021). A driving simulator study for assessing the potential of augmented reality technology to improve the safety of passing maneuvers. 2021 7th International Conference on Models and Technologies for Intelligent Transportation Systems (MT-ITS), 1–6. https://doi.org/10.1109/MT-ITS49943.2021.9529339

Cohen, J. (2013). Statistical power analysis for the behavioral sciences (0 ed.). Routledge. https://doi.org/10.4324/9780203771587

De Koning, B. B., Tabbers, H. K., Rikers, R. M. J. P., & Paas, F. (2010). Attention guidance in learning from a complex animation: Seeing is understanding? Learning and Instruction, 20(2), 111–122. https://doi.org/10.1016/j.learninstruc.2009.02.010

Demir, M., Cohen, M., Johnson, C. J., Chiou, E. K., & Cooke, N. J. (2023). Exploration of the Impact of Interpersonal Communication and Coordination Dynamics on Team Effectiveness in Human-Machine Teams. International Journal of Human–Computer Interaction, 39(9), 1841–1855. https://doi.org/10.1080/10447318.2022.2143004

Endsley, M. (1995). A taxonomy of situation awareness errors. Human Factors in Aviation Operations: Proceedings of the 21st Conference of the European Association for Aviation Psychology (EAAP), 1.

Endsley, M. (2001). Designing for situation awareness in complex systems. Proceedings of the Second International Workshop on Symbiosis of Humans, Artifacts and Environment, 1–14.

Endsley, M., Bolte, B., & Jones, D. (2003). SA Demons: The Enemies of Situation Awareness. In Designing for situation awareness: An approach to user-centered design (1st ed.). CRC Press.

Faul, F., Erdfelder, E., Lang, A.-G., & Buchner, A. (2007). G*Power 3: A flexible statistical power analysis program for the social, behavioral, and biomedical sciences. Behavior Research Methods, 39(2), 175–191. https://doi.org/10.3758/BF03193146

Fletcher, J. D. (2009). Education and training technology in the military. Science, 323(5910), 72–75. https://doi.org/10.1126/science.1167778

Fontes, R., & Kamminga, J. (2023, March 24). Ukraine a living lab for AI warfare. National Defense. https://www.nationaldefensemagazine.org/articles/2023/3/24/ukraine-a-living-lab-for-ai-warfare

Fortmann, F., & Lüdtke, A. (2013). An intelligent SA-adaptive interface to aid supervisory control of a UAV swarm. 2013 11th IEEE International Conference on Industrial Informatics (INDIN), 768–773. https://doi.org/10.1109/INDIN.2013.6622981

Fortmann, F., & Mengeringhausen, T. (2014). Development and evaluation of an assistant system to aid monitoring behavior during multi-UAV supervisory control: Experiences from the D3CoS project. Proceedings of the 2014 European Conference on Cognitive Ergonomics, 1–8. https://doi.org/10.1145/2637248.2637257

Frame, M., Maresca, A., Christensen-Salem, A., & Patterson, R. (2022). Evaluation of simulated recognition aids for human sensemaking in applied surveillance scenarios. Human Factors. https://doi.org/10.1177/00187208221120461

Grant, E. R., & Spivey, M. J. (2003). Eye movements and problem solving: Guiding attention guides thought. Psychological Science, 14(5), 462–466. https://doi.org/10.1111/1467-9280.02454

Hopp, P., Smith, C., Clegg, B., & Heggestad, E. (2005). Interruption management: The use of attention-directing tactile cues. Human Factors, 47(1), 1–11. https://doi.org/10.1518/0018720053653884

Humr, S. A., Canan, M., & Demir, M. (2023). Temporal Evolution of Trust in Artificial Intelligence-Supported Decision-Making. Proceedings of the Human Factors and Ergonomics Society Annual Meeting, 21695067231193672. https://doi.org/10.1177/21695067231193672

Humr, S., Canan, M., & Demir, M. (2025). A Quantum Probability Approach to Improving Human–AI Decision Making. Entropy, 27(2), 152. https://doi.org/10.3390/e27020152

Joint Staff J-7. (2011). Commander's handbook for assessment planning and execution. Department of Defense.

Krejtz, K., Duchowski, A. T., Niedzielska, A., Biele, C., & Krejtz, I. (2018). Eye tracking cognitive load using pupil diameter and microsaccades with fixed gaze. PLOS ONE, 13(9), e0203629. https://doi.org/10.1371/journal.pone.0203629

Lewandowska, A., Dzisko, M., & Jankowski, J. (2022). Investigation the role of contrast on habituation and sensitisation effects in peripheral areas of graphical user interfaces. Scientific Reports, 12(1). https://doi.org/10.1038/s41598-022-16284-2

Mack, A. (2003). Inattentional Blindness: Looking Without Seeing. Current Directions in Psychological Science, 12(5), 180–184. https://doi.org/10.1111/1467-8721.01256

Mack, A., & Rock, I. (1998). Inattentional blindness: Perception without attention. In Visual Attention.

Monk, C. A., Trafton, J. G., & Boehm-Davis, D. A. (2008). The effect of interruption duration and demand on resuming suspended goals. Journal of Experimental Psychology: Applied, 14(4), 299–313. https://doi.org/10.1037/a0014402

Neisser, U. (1979). The Control of Information Pickup in Selective Looking. In Perception and Its Development (1st Edition).

Nonino, E., Gisler, J., Holzwarth, V., Hirt, C., & Kunz, A. (2021). Subtle attention guidance for real walking in virtual environments (WOS:000758410300058). 310–315. https://doi.org/10.1109/ISMAR-Adjunct54149.2021.00070

Ratwani, R. M., McCurry, J. M., & Trafton, J. G. (2008). Predicting postcompletion errors using eye movements. Proceedings of the SIGCHI Conference on Human Factors in Computing Systems, 539–542. https://doi.org/10.1145/1357054.1357141

Ratwani, R. M., & Trafton, J. G. (2011). A real-time eye tracking system for predicting and preventing postcompletion errors. Human–Computer Interaction, 26(3), 205–245. https://doi.org/10.1080/07370024.2011.601692

Rensink, R. A., O'Regan, J. K., & Clark, J. J. (1997). To See or not to See: The Need for Attention to Perceive Changes in Scenes. Psychological Science, 8(5), 368–373. https://doi.org/10.1111/j.1467-9280.1997.tb00427.x

Rickli, J.-M., & Mantellassi, F. (2024). GCSP Publication | The War in Ukraine: Reality Check for Emerging Technologies and the Future of Warfare. Centre for Security Policy. https://www.gcsp.ch/publications/war-ukraine-reality-check-emerging-technologies-and-future-warfare

Roads, B., Mozer, M., & Busey, T. (2016). Using Highlighting to Train Attentional Expertise. PLOS ONE, 11(1). https://doi.org/10.1371/journal.pone.0146266

Rogers, T. (2023). Human information processing under stress. Journal of Emergency Management, 21(2), 141–154. https://doi.org/10.5055/jem.0756

Sasangohar, F., Scott, S. D., & Cummings, M. L. (2014). Supervisory-level interruption recovery in time-critical control tasks. Applied Ergonomics, 45(4), 1148–1156. https://doi.org/10.1016/j.apergo.2014.02.005

Savick, D. S., Elliott, L. R., Zubal, O., & Stachowiak, C. (2008). The effect of audio and tactile cues on soldier decision making and navigation in complex simulation scenarios (No. ARL-TR-4413). U.S. Army Research Laboratory.

Scott, S. D., Sasangohar, F., & Cummings, M. L. (2009). Investigating supervisory-level activity awareness displays for command and control operations. Proceedings of HSIS 2009: Human Systems Integration Symposium. Human Systems Integration Symposium, Annapolis, MD, USA.

Simons, D. J., & Chabris, C. F. (1999). Gorillas in Our Midst: Sustained Inattentional Blindness for Dynamic Events. Perception, 28(9), 1059–1074. https://doi.org/10.1068/p281059

Simons, D. J., & Levin, D. T. (1997). Change blindness. Trends in Cognitive Sciences, 1(7), 261–267. https://doi.org/10.1016/S1364-6613(97)01080-2

Sklar, A. E., & Sarter, N. B. (1999). Good vibrations: Tactile feedback in support of attention allocation and human-automation coordination in event-driven domains. Human Factors, 41(4), 543–552. https://doi.org/10.1518/001872099779656716

Stahl, P., Donmez, B., & Jamieson, G. (2016). Supporting anticipation in driving through attentional and interpretational in-vehicle displays. Accident Analysis and Prevention, 91, 103–113. https://doi.org/10.1016/j.aap.2016.02.030

Takahashi, H. (2017). Visual cue in the peripheral vision field for a driving support system. Journal of Advanced Computational Intelligence and Intelligent Informatics, 21(3), 543–558. https://doi.org/10.20965/jaciii.2017.p0543

Tasse, D., Ankolekar, A., & Hailpern, J. (2016). Getting users' attention in web apps in likable, minimally annoying ways. Proceedings of the 2016 CHI Conference on Human Factors in Computing Systems, 3324–3334. https://doi.org/10.1145/2858036.2858174

Tobii AB. (2019). Tobii Pro Lab (Version 1.123) [English]. Tobii AB.

Tobii AB. (2023, September). Tobii Pro Nano [User Manual v2.5]. Tobii Pro Nano. https://go.tobii.com/tobii-pro-nano-user-manual

Trafton, J. G., Altmann, E. M., & Brock, D. P. (2005). Huh, what was I doing? How people use environmental cues after an interruption. Proceedings of the Human Factors and Ergonomics Society Annual Meeting, 49(3), 468–472. https://doi.org/10.1177/154193120504900354

Tsotsos, J. K., Culhane, S. M., Kei Wai, W. Y., Lai, Y., Davis, N., & Nuflo, F. (1995). Modeling visual attention via selective tuning. Artificial Intelligence, 78(1–2), 507–545. https://doi.org/10.1016/0004-3702(95)00025-9

Van Der Wel, P., & Van Steenbergen, H. (2018). Pupil dilation as an index of effort in cognitive control tasks: A review. Psychonomic Bulletin & Review, 25(6), 2005–2015. https://doi.org/10.3758/s13423-018-1432-y

Van Gerven, P. W. M., Paas, F., Van Merriënboer, J. J. G., & Schmidt, H. G. (2004). Memory load and the cognitive pupillary response in aging. Psychophysiology, 41(2), 167–174. https://doi.org/10.1111/j.1469-8986.2003.00148.x

Vlassova, A., Donkin, C., & Pearson, J. (2014). Unconscious information changes decision accuracy but not confidence. Proceedings of the National Academy of Sciences, 111(45), 16214–16218. https://doi.org/10.1073/pnas.1403619111

Ward, M., & Helton, W. (2022). More or less? Improving monocular head mounted display assisted visual search by reducing guidance precision. Applied Ergonomics, 102. https://doi.org/10.1016/j.apergo.2022.103720

Wickens, C. D. (1976). The effects of divided attention on information processing in manual tracking. Journal of Experimental Psychology: Human Perception and Performance, 2(1), 1–13. https://doi.org/10.1037/0096-1523.2.1.1

Wickens, C. D. (2015). Noticing Events in the Visual Workplace: The SEEV and NSEEV Models. In R. R. Hoffman, P. A. Hancock, M. W. Scerbo, R. Parasuraman, & J. L. Szalma (Eds.), The Cambridge Handbook of Applied Perception Research (1st ed., pp. 749–768). Cambridge University Press. https://doi.org/10.1017/CBO9780511973017.046

Wickens, C. D., & Carswell, C. M. (1995). The proximity compatibility principle: Its psychological foundation and relevance to display design. Human Factors: The Journal of the Human Factors and Ergonomics Society, 37(3), 473–494. https://doi.org/10.1518/001872095779049408

**APPENDIX**

Semi-structured post-experiment interview questions. There were seven pre-designated questions, but follow-up questions were asked when necessary.

1. What did you think about the experiment? What was your experience like?
2. What was the most helpful/useful feature on the displays when gaining situational awareness, processing information, and making decisions?
3. Did you notice anything different between the two trials? If yes, what was different?
4. Did you find the interruption tasks distracting enough to take your mind off the mission? Why?
5. Did you find the interruption tasks long enough to take your mind off the mission? Why?
6. Did you find the task or experimental environment realistic enough? Do you have any recommendations to make it more realistic?
7. Do you have any other comments or suggestions about the usability of the interfaces or the mission task?